# Pulli Kolam: A Traditional South Indian Craft Practice for Representing Data

**Shri Harini Ramesh**
University of Calgary
Calgary, Canada
Shri.ramesh@ucalgary.ca

**Fateme Rajabiyazdi**
University of Calgary
Calgary, Canada
fateme.rajabiyazdi@ucalgary.ca

## INTRODUCTION

In this statement, we introduce Pulli Kolam, a traditional South Indian craft, and envision how this craft can be used to represent data physically. In South Indian households, particularly in Tamil Nadu, creating Kolam is a daily ritual aimed at welcoming prosperity and well-being into the home. Kolam is performed primarily by women at home entrances before sunrise. Kolam is associated with cultural and physical well-being. Culturally, it welcomes Goddess Lakshmi (the deity of wealth and prosperity), and the use of rice flour serves as an offering to ants and other small creatures, embodying the principle of coexistence with nature [1]. Physically, creating Kolam is an embodied practice [2]. To reach the ground and follow the pattern, the maker must sustain a bent-waist or deep-squatting posture for extended periods (see Figure 1).

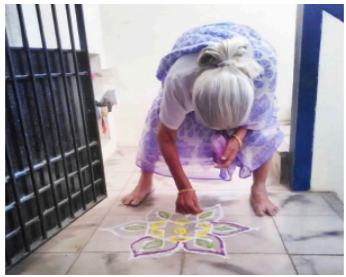

Figure 1. A woman in a South Indian household is in a bending posture while making a kolam [4].

This daily practice engages the core, stretches muscles, and strengthens flexibility, making the creation of Kolam a structured, physically mindful activity as much as a visual craft [3].

The Kolam process begins with preparing the surface (that is, stone or packed dirt) by sweeping, washing, and sometimes coating with cow dung for a smooth, dark canvas. Using rice flour held in a bowl or coconut shell, the maker uses a pinch-and-release technique with their fingers to first lay down a dot grid (pulli) (See Figure 2(a)), then draws continuous geometric lines looping around these dots, forming complex, symmetrical patterns (see Figure 2(b)).

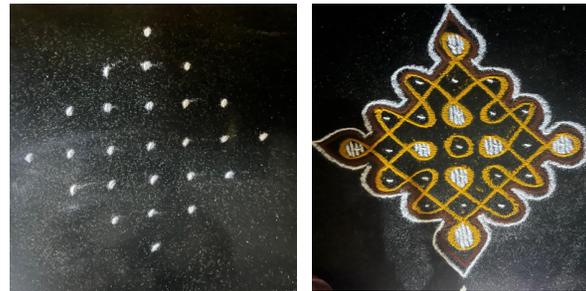

Figure 2. (a) A dot grid laid out on the ground, serving as the basis for the kolam. (b) Continuous lines drawn around the dots, forming a kolam pattern.

Pulli Kolam follows a set of rules that create the structure of the Kolam while allowing creativity [5]: (1) **The obstacle rule:** Once the dot grid is drawn, dots act as rigid obstacles. Lines must never touch a dot; they must be drawn around it. (2) **The continuity rule:** Lines must form a single, continuous path that loops back to its starting point without breaking, a Eulerian Circuit in graph theory. (3) **The completeness rule**: Every dot must be enclosed by the line, and patterns must have reflective or rotational symmetry. No dot can be left stranded outside the pattern. (4) **The smoothness rule**: Lines must curve gently around dots without sharp corners, creating smooth, fluid arcs. (5) **The diagonal rule**: Lines can be straight only when travelling diagonally at 45 degrees through the spaces between dots.

## DATA PHYSICALIZATION THROUGH KOLAM

Pulli Kolam's creation rules and embodied practices could offer multiple opportunities for representing data physically.

**1. Representing data through the dots (Pulli):**
The dot grid is the structural base of Kolam and can be used to represent data by varying dot size (see Figure 3). In Kolam practice, dot size is controlled by finger pressure when releasing rice flour, resulting in dots of different sizes. By adjusting this pressure, dots can be made larger or smaller, thereby enabling systematic variation across the grid.

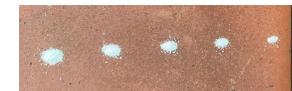

Figure 3. Various dot sizes were created using rice flour.

**2. Representing data through patterns:**
Multiple Kolam patterns can be generated from the same spatial arrangement of dots. Using an identical grid, for example, a 3×3 dot layout, different line paths can be drawn to produce distinct patterns (see Figure 4 (a)). Each resulting pattern can be treated as a separate representation while preserving the underlying grid structure. This can support variation at the pattern level without changing dot placement. This could allow mapping different categorical attributes such as mood. For instance, one pattern could represent a calm day, another an energetic day, and another a stressful day.

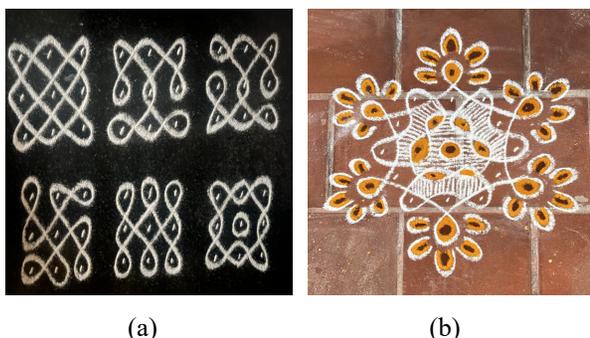

(a)            (b)

Figure 4. (a) Kolam patterns created using the same 3×3 dot grid. (b) Kolam with horizontal and diagonal fill patterns.

### 3. Representing data through fill patterns:
Once the Kolam line forms enclosed regions, these areas can be treated as containers for representing data. Data can be represented by varying fill presence, fill pattern, and pattern density. Regions can be left empty or filled to indicate differences in data. Data can also be represented using different fill patterns (see Figure 4(b)), such as horizontal lines, vertical lines, diagonal lines, checkerboard patterns, dots, or concentric shapes that follow the boundary of the enclosed region. Pattern density, ranging from tightly packed to sparsely spaced fills, can also be used to represent data.

### 4. Representing data through the line:
The continuous line drawn around the dots can be used to represent data by varying the line type (see Figure 5(a)). Different types of lines can be created through specific finger positions when holding the powder. A single continuous line is created by pinching the powder between the thumb and forefinger. Parallel double lines are formed by holding the powder between the thumb, forefinger, and middle finger with spacing between the fingers. Double lines with a filled center are created by holding the same fingers tightly together, allowing powder to flow into the center gap.

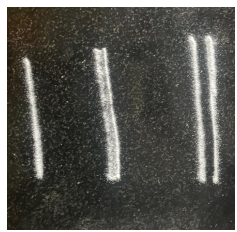

Figure 5. Three types of lines created using different finger positions.

### 5. Representing data through color:
Color can also be used to represent data in Kolam. Traditional Kolam materials include white rice flour as the base medium. When color is introduced, commonly used materials include yellow turmeric, red kumkum, and reddish-brown pigment derived from laterite soil (see Figure 5(b)). These colors may be used individually or mixed to produce intermediate shades. While traditional materials are typically used, non-traditional pigments can also be introduced by mixing artificial pigments with white rice flour. This extends the available color range and allows for more granular variation when mapping data.

### USE CASE
To iillustrate how the mapping strategies described above can be applied in practice, we present an illustrative scenario centered on Lakshmi, a software engineer living in Chennai, Tamil Nadu, India, who has been creating Kolam every morning for over fifteen years. Creating Kolam is the first activity of her day, a practice she learned from her grandmother and has maintained consistently over the years. Lakshmi chose to use this daily activity as a moment to reflect on the previous day and to represent aspects of her personal well-being. She identified four data attributes that were meaningful to her: hours of sleep, overall energy level, mood, and physical activity.

Lakshmi developed a mapping scheme through gradual experimentation, ensuring that each mapping fit naturally within her existing Kolam-making process (see Figure 6). She represented sleep through the size of the dots (pulli). Nights with a longer duration of sleep are represented by larger dots, while short nights result in smaller dots. Energy level is represented through line type. High-energy days are represented using parallel double lines, moderate-energy days with double lines and a filled center, and low-energy days with a single continuous line. She represented mood through color using materials already present in the household: calm or emotionally balanced days are drawn with white rice flour, positive days with yellow turmeric, and stressful days with red kumkum. Physical activity is represented through fill patterns within enclosed loops. If some form of physical activity is completed, the loops are filled; if not, they are left empty. Different activities are distinguished through fill patterns, such as vertical lines for strength training and horizontal lines for yoga.

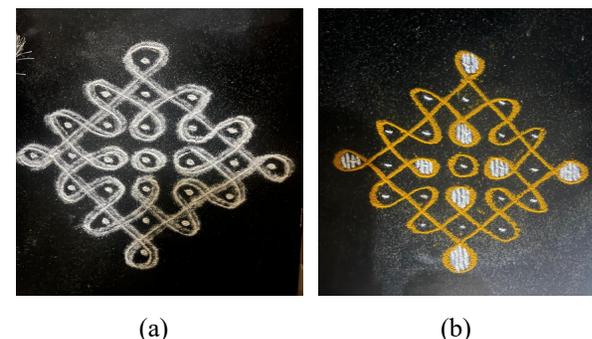

(a)            (b)

Figure 6. Lakshmi's kolam on two consecutive days. Day 1: large dots, double lines with filled centers, white rice flour, and empty loops. Day 2: small dots, single lines, yellow turmeric, and vertical-fill loops.

## CONCLUSION AND FUTURE WORK

In this statement, we introduced Kolam, a traditional craft practice of South India, and envisioned how it can be used to represent data physically. We demonstrated five mapping strategies (dots, patterns, fills, lines, and color) and explained how these mapping strategies may enable data representation while preserving the structural rules and embodied practices of traditional Kolam-making.

As kolam patterns are cleaned daily to make a new design, future research can explore adaptations such as creating artifacts for people to track and share their data with others in ways that can be preserved and displayed. For example, Nagata et al. created small 3D cubes with each face of the cube representing a part of a kolam [6], which can be arranged together to create various kolam patterns (see Figure 7(a)). People can physically rearrange the cubes daily, creating new patterns for representing data. Another example could be daily Kolam patterns stitched onto fabric patches using different stitch sizes, thread weights, and colors to represent data variables (see Figure 7(b)). Over time, these patches could be assembled into decorative household items such as cushion covers [7], table runners, or wall hangings, transforming the data archive into functional home decor.

By reimagining Kolam patterns through different materials and forms, we can overcome the ephemeral nature of traditional Kolam practice when long-term tracking is needed while preserving the hands-on, embodied interaction central to Kolam-making craft. Ultimately, Kolam can serve as a meaningful medium for physical data representation, where the daily ritual of making it simultaneously an act of tracking and reflecting.

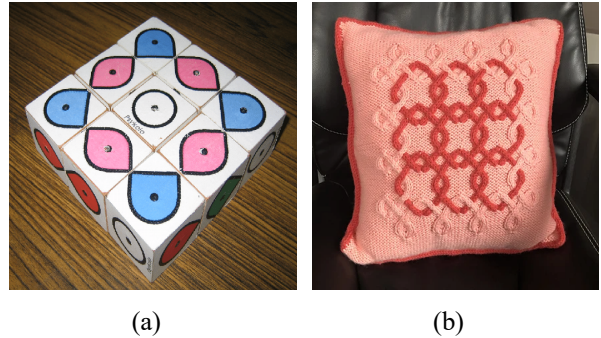

(a)      (b)

Figure 7. Examples of Kolam adaptations (a) Modular cubes arranged in grid patterns [6]. (b) Embroidered Kolam patterns [7].

## ACKNOWLEDGEMENTS
We would like to thank Nithyajayalakshmi Swaminathan (the first author's mother) for creating the kolam patterns and generously providing the images used in this work.